# Tailoring phase slip events through magnetic doping in superconductor-ferromagnet composite films

*Ambika Bawa, Rajveer Jha & Sangeeta Sahoo\**

*Quantum Phenomena & Applications, CSIR- National Physical Laboratory, Council of Scientific and Industrial Research, Dr. K. S. Krishnan Marg, New Delhi, India- 110012*

*\*Correspondence and requests for materials should be addressed to S.S. (sahoos@nplindia.org)*

The interplay between superconductivity (SC) and ferromagnetism (FM) when embedded together has attracted unprecedented research interest due to very rare coexistence of these two phenomena. The focus has been mainly put into the proximity induced effects like, coexistence of magnetism and superconductivity, higher critical current, triplet superconductivity etc. However, very little attention has been paid experimentally to the role of magnetic constituent on triggering phase slip processes in the composite films (CFs). We demonstrate that less than 1 at. % of magnetic contribution in the CFs can initiate phase slip events efficiently. Due to advanced state-of-the-art fabrication techniques, phase slip based studies have been concentrated mainly on superconducting nanostructures. Here, we employ wide mesoscopic NbGd based CFs to study the phase slip processes. Low temperature current-voltage characteristics (IVCs) of CFs show stair-like features originated through phase slip events and are absent in pure SC films. Depending on the bias current and temperature, distinct regions, dominated by Abrikosov type vortex-antivortex (v-av) pairs and phase slip events, are observed. The results presented here open a new way to study the phase slip mechanism, its interaction with v-av pairs in two dimensions and hence can be useful for future photonic and metrological applications

One of the main mechanisms for dissipation in current carrying thin films of superconductor is phase slip events. During a phase slip event, phase difference between two parts of a one-dimensional (1D) superconducting wire changes by $2\pi$ while the superconducting order parameter fluctuates to zero leading to the nucleation of a phase slip center (PSC). On the fundamental level, the activation mechanisms of phase slips are of thermal origin at temperature close to the critical temperature (Tc) and far below the Tc, quantum fluctuations[1] take charge of the events. On the technological perspectives, a phase slip junction[2] where coherent phase slippage takes place may be considered as the dual counterpart of the Josephson junction with phase difference replaced by charge[3-5]. Thus like Josephson voltage standard, a phase slip junction can lead to the realization of the fundamental current standard[6].

Recently, it has been experimentally demonstrated that a disordered superconducting wire can lead to coherent quantum phase slip (CQPS)[7] while theoretical studies revealed that a weak inhomogeneity in superconducting wire can be treated as a weak link having less resistance than the overall normal state resistance and can be favorable for localizing a phase slip junction[5]. Independent studies suggested that incorporation of magnetic impurity into a superconducting matrix is able to initiate, affect the phase-slip events and hence can be used in a controlled manner to preserve the coherence in nanostructures and also to study the dissipation effects in superconducting materials[8]. Besides, a magnetic dot can act as a very effective phase slip pinning center and thus directly influence the dissipation dynamics in the superconducting condensate[9].

Similar to PSCs in 1D superconducting nanowires, in wide two-dimensional (2D) mesoscopic superconducting thin films, phase slip lines (PSLs) are formed which can act effectively as Josephson weak-links[10]. PSCs or PSLs appear as voltage steps-like features in both IVCs[11,12] and in voltage vs. temperature characteristics[13] and IVCs display a wide transition region between the appearance of first non-zero voltage and the completely ohmic behavior representing its normal (NM) state[8]. Apart from PSLs, for clean superconducting films in 2D, flux free motion of Abrikosov vortex-antivortex (v-av) pairs takes a vital role in the dissipation mechanism under the application of external bias current. On the other hand, , the v-av pairs creation and their trajectories get influenced by the impurities in the dirty limit particularly with magnetic impurities[14-16] . For example, in planar FM/SC hybrid structures, v-av pairs have been demonstrated[17]. Thus incorporation of magnetic dopant molecules in 2D mesoscopic superconducting systems can serve as a model system to study the interplay between the phase slip process and the Abrikosov vortex lattice since the latter can contribute significantly to the dissipation[18]. Here, we aim to address the impact of magnetic doping on the initiation and modulation of phase slip processes for the SC to NM state transition. We consider Nb (SC) and Gd (FM) based 2D CFs with varying Gd concentration. Gd concentration is varied within a narrow range between 0 to1 at. % so that any reasonable stray field arising from the magnetic GD part can be ignored. We study the evolution of IVCs with temperature and Gd concentration. Hysteretic IVCs are observed for all the samples including pure Nb films and have been attributed to the self-heating under the current biasing condition. Contrary to the Nb samples, stair-like structures indicating the possible formation of phase-slip lines are observed in the IVCs for all 4 NbGd samples. Absence of these features in the IVCs for pure Nb films indicates the phase slippage is triggered by magnetic Gd counterpart in the CFs. Below the switching current, smooth variation in voltage with current indicates a possible flow of v-av pairs. The measurements reveal a strong dependence of the critical current on the Gd concentration and we find a relatively faster decay in the critical current with temperature for higher Gd doping.

## Results:

The surface morphologies of thin films are characterized by an atomic force microscopy (AFM) in non-contact mode. We have characterized both Nb and NbGd films having different Gd concentration. No qualitative differences have been observed for Nb films and NbGd CFs with Gd up to 5 at. %. The topography image for one representative sample is shown in Figure-1. The granular structure is very clear in the AFM image of 100 nm thick Nb thin film deposited on Si substrate with 300 nm thick thermally grown SiO$_2$ on top of it. The films were grown at room temperature using an UHV magnetron sputtering system. Closely packed nanostructured grains of 20 to 30 nm in size appear in the AFM morphology and hence they can be considered as weakly coupled disordered Josephson junction network[19].

Multi-terminal devices based on NbGd thin films with different Gd concentration, $c$, have been fabricated for the transport studies. A typical device structure is shown in the inset of Figure 2(a). The dark vertical 100 μm wide line represents the NbGd channel connected with two current leads (top and Bottom leads) and two voltage leads (two outer metallic fingers). The resistance vs. temperature (R-T) and the resistance vs. magnetic field (R-H) measurements, in four probe geometry, are carried out in current biasing mode with a low frequency (17 Hz) ac excitation of 10 μA. The transport properties were measured in a Physical Properties Measurements System (PPMS) from Quantum Design Inc.

We display a set of R-T characteristics for 4 NbGd devices and one for Nb device in Figure 2. The device S#1 corresponds to only Nb film with thickness about (50 ± 5) nm. The samples S#2 to S#5 are NbGd devices of thickness about (110 ± 10) nm with varying $c$ measured by energy dispersive spectroscopy (EDS). The values of $c$ for individual devices are mentioned in Figure 2(a) along with their R-T characteristics. While cooling down the sample from room temperature, a sharp NM to SC phase transition occurs at 8.75 K for the Nb device, S#1 (black points). The transition width, the difference in the temperature for 90% and 10% of the normal state resistance, for the Nb device is about 80 mK indicating a good quality film having resistivity much less than the NbGd devices shown in Figure 2(a). With increasing $c$, the transition shifts towards lower temperature while getting wider. For example, samples, S#4 and S#5, show broad transitions with intermediate features, while S#2 and S#3 with lesser Gd part show much sharper transition. The wide transitions for the formers contain a relatively sharp drop in resistivity followed by a kink/tail type of feature indicating much weaker dependence of R on T. Similar features have been reported for superconducting nanowires[20], constrictions[21] and high-T$_c$ micro-bridges[22] where the wide tail parts are attributed to occurrence of possible phase slip events. The initial sharp drop in resistance and the wide transition tail can be interpreted as the superconducting transition and thermally activated phase slip events respectively[23]. Further, it is evident that magnetic impurities play an important role for the transition temperature as well as the transition width and the dynamics of the NM-SC quantum phase transition.

The R-T curves, shown in Figure 2(a), are plotted in semi-logarithmic scale in Figure 2(b). For sample S#5, a plateau type feature, marked with dotted circular region, with resistance less than 0.03% of its normal state value appears. The plateau remains almost constant over an extent of 0.5 K indicating phase slippage as the possible reason for the wide feature appeared in the R-T. The stability, reproducibility, and the evolution of the resistance plateau for S#5 under an external magnetic field applied normal to the sample plane are examined in Figure 3. Figure 3(a) presents a set of R-T curves measured under different field. As expected, the transition moves to a lower temperature with increasing field and superconductivity is lost at about 500 mT. The zero-field plateau, highlighted by the black dashed arrow, appears also at 50 mT (the red curve). At 100 mT, the transition gets broadened and the feature disappears from the measureable temperature range. Besides, the R-H measurements at constant temperature firmly support the existence of this intermediate state as shown in Figure 3(b). From 2 K to 2.5 K, we observe a similar type of plateau in the resistance with field before reaching to zero resistance. The extent of this region in magnetic field decreases with increasing temperature and disappears above 3 K. The plateau region extends up to about 75 mT at 2 K. Thus, the distinct feature represented by the plateau in zero-field R-T curve for sample S#5 is retrieved at moderately low field demonstrating its stability and reproducibility under external field.

To further investigate the phase slip mechanisms triggered by the magnetic impurity, we have measured isothermal IVCs in 4-probe geometry with current biasing mode using the PPMS setup and here we present the measured IVCs for some of the devices shown in Figure 2. First, we exhibit the IVC isotherms for pure Nb sample, S#1, in Figure 4. Arrows indicate the current sweeping direction on which IVCs appear to be dependent strongly. The individual IVC isotherms are separately displayed along the z-axis using a 3-dimensional (3D) frame in Figure 4(b) for clarity. The hysteretic IVCs are evident. The temperature dependent hysteresis decreases with increasing temperature and eventually it disappears at critical temperature ($T_c$).

Critical current ($I_{c0}$) and the retrapping current ($I_r$) are extracted from the IVCs and are plotted in the inset of Figure 4(b). The critical current is defined as the current related to the onset of a non-zero voltage from its SC state. For the direction of increasing bias current (up sweep), a finite voltage ($V \neq 0$) starts to appear at the critical current, $I_{c1}^{up}$ and for decreasing current direction (down sweep), the critical current, $I_{c1}^{dn}$, is determined by the current above which a finite voltage exists. $I_r$ is represented by the current where the onset of superconducting transition from its resistive state occurs during the down sweep. The arrows in Figure 4(b) describe the position of the afore-mentioned characteristic current values. The transport measurements are performed on 3 Nb devices and for all of them, the switching current related to a jump from superconducting to resistive state merge with their critical current even at temperature close to its $T_c$ without showing any gradual dissipation process.

In the inset of Figure 4(b), the temperature dependent critical current, $I_{c1}^{up}(T)$, appears at much higher current than $I_r(T)$ at temperatures far below $T_c$. As we approach towards the $T_c$, $I_{c1}^{up}(T)$ and $I_r(T)$ merge and the hysteresis disappears. We define a characteristic threshold temperature, $T_h$, as the temperature where $I_{c1}^{up}(T)$ and $I_r(T)$ meet and is shown by the arrow in the inset of Figure 4(b). Above $T_h$, hysteresis in the IVCs are expected to disappear[24]. $I_{c1}^{dn}(T)$ and $I_r(T)$ emerge very close to each other indicating a very sharp transition from a resistive to superconductive state. For superconducting nanostructured systems, especially for WLs, hysteretic IVCs are very common phenomena and the hysteresis appear due to the Joule heating, which raises the temperature locally and the increased effective temperature reduces the critical current [25][26]. Thus, the hysteresis can be explained by the self-heating in Josephson like WLs[27] formed by the closely packed nanostructured grains as evident in Figure 1.

Now the question arises that how different are the IVCs for NbGd devices than that from Nb samples. We selectively present isothermal IVCs for two NbGd samples, S#3 and S#5, in Figures 5 and 6, respectively. For S#3 IVCs are plotted separately for each temperature along the Z-axis in Figure 5 (a). Similar to the Nb devices, the IVCs appear to be hysteretic with the direction of current sweep and the hysteresis gets reduced with increasing temperature. More interestingly, from 6.0 K onwards IVCs become significantly different than a single sharp transition from superconducting to resistive (resistive to superconducting) state for up (down) sweeps respectively. Four representative IVC isotherms, shifted from subsequent one by 50 mV for clarity, are presented separately in Figure 5(b). Steps-like features start to appear first for the down sweep at 6.0 K [Figure 5 (a) & (b)]. At 6.1 K, both up and down sweeps start to display multiple steps during the transition (green points in Figure 5b). With increasing temperature, the steps in the IVCs get extended and eventually merge with the neighboring ones. At temperatures close to $T_c$, the transition gets smoother and the hysteresis starts to disappear [Figure 5 (a)]. Extrapolations, shown by the dotted lines, from the voltage steps converge at a single point on the current axis indicating the existence of an excess current. Further, the dynamic resistance, given by the slope of a step in a IVC increases with the increase in step number. Hence, it is evident that these step-like features originate through phase slip events[10,11,28]. Moreover, appearance of these features in NbGd devices in contrast to Nb devices implies the magnetic dopant triggered phase slip events in the NbGd films.

Characteristic current values are indicated by the arrows in Figure 5(b). $I_{c0}$ is the critical current at which a non-zero voltage (V ≠ 0) starts to appear for superconducting to resistive transition in up sweep and is shown for the IVC at 6.3 K. $I_{c1}^{up}$ is defined as the first switching current where a voltage jump occurs from the superconducting state to the first resistive state for up sweep. During the down sweep, onset of the transition from normal state to the superconducting and/or the intermediate resistive state is defined as the retrapping current, $I_r$ while $I_{c1}^{dn}$ indicates the current where voltage becomes zero for the transition from a resistive state to the superconducting state. The temperature dependence of these characteristic currents, $I_{c0}(T)$, $I_{c1}^{up}(T)$, $I_{c1}^{dn}(T)$, and $I_r(T)$, are displayed in Figure 5 (c). The difference between $I_{c1}^{dn}$

(T) (blue points) and $I_r$(T) (red points) is clearly evident in contrast to that for Nb and it measures the transition width containing the information of the dissipation dynamics for down sweep. The difference between $I_{c0}^{up}$(T) (black points) and $I_{c1}^{up}$(T) (green points) relates to the dissipation dynamics in superconducting state. For example, up to 5.5 K both the curves merge onto each other indicating a direct transition from superconducting to resistive state at the critical current, while above 5.5 K the curves start to separate and residual voltage tails start to appear just before the switching. In Figure 5 (b) at 6.3 K, a smoothly varying non-zero voltage is clearly observed. The voltage tails may appear due to either the flux flow of Abrikosov vortices from the edges of the sample in 2D [10] or by the spontaneous formation of v-av pairs due to the magnetic granules[29]. Thus, the region bounded by $I_{c0}^{up}$(T) and $I_{c1}^{up}$(T) provides a regime where mainly v-av pairs participate into the dissipation process[10]. The threshold temperatures, $T_h$ and $T_h^{c1}$, are shown in Figure 5 (c) as the intersection points of $I_r$(T) to the $I_{c0}^{up}$(T) and $I_{c1}^{up}$(T) curves respectively. Contrary to the Nb samples, the hysteresis for this NbGd device does not disappear above $T_h$ and is present even above $T_h^{c1}$. Other noticeable feature is that $I_r$ becomes higher than the critical current above the threshold temperature. It has been shown that superconducting WLs might be in the resistive state while its temperature stays below $T_c$ and in this case at temperature higher than $T_h$, a regime can be attained where $I_r$ exceeds $I_{c0}$ [24]. At this situation, SC-NM transition occurs only when the bias current is larger than $I_r$.

A set of isothermal IVCs for NbGd device S#5 are displayed in Figure 6(a). Similar to S#3, a set of clear distinct voltage steps appear in the IVCs. Three main voltage steps related to three intermediate resistive states are observed in the IVCs at lower temperature and the corresponding switching currents are specified by the arrows. The steps move towards the lower current with increasing temperature. Interestingly, for different isothermal IVCs any individual step, for example the 1$^{st}$ resistive state after the first switching from the superconducting state, follow the same slope as shown by the dotted line in Figure 6 (a). Further, the slopes for the major three steps are extended with the dotted lines and their convergence to a single point is very clear from Figure 6 (a). The convergence of the slopes and the increment in the slope for higher order steps once more demonstrate the evidence towards the phase slippage in these NbGd samples[10].

As for the isotherms falling close to each other in Figure 6(a), we present the same IVCs shifted by 35 mV from the preceding one in Figure 6(b) for clearness. In order to study the evolution of voltage steps with temperature, we have marked three regions representing three intermediate resistive states with dotted lines along their numberings for the IVC measured at 2.0 K in Figure 6(b). The area of individual regions gets reduced and overall hysteretic behavior of IVCs follow a diminishing trend with increasing temperature. For example, as the bath temperature increases from 2.0 K to 2.6 K, the $1^{st}$ zone close to the superconducting state starts to decrease in size and disappears at 2.6 K. At 2.7 K, the first switching starts from $2^{nd}$ –zone; thus with increasing temperature number of PSLs characterized by the voltage steps gets reduced. With increasing temperature, for a given value of current density, number of PSLs decreases since at high temperature effective size of the sample in units of the coherence length and the penetration depth gets reduced to accommodate more PSLs[13].

The temperature dependence of characteristic current values, $I_{c0}^{up}$, $I_{c1}^{up}$, $I_{c1}^{dn}$, $I_r$, defined similarly as done for S#3 are shown in Figure 6(c). Below 2.3 K, $I_{c0}^{up}$ and $I_{c1}^{up}$ remain unaltered and from 2.3 K onwards they start to deviate and follow separate path. The significant difference between $I_{c1}^{dn}(T)$ and $I_r(T)$ indicates the wide nature of the transition during the returning sweep as has been reflected also in the respective R-T measurements. Further, the threshold temperature $T_h$ is closed to 2.0 K which is much lower than S#3 and S#1. Yet again above $T_h$, $I_r$ becomes higher than $I_{c0}$ for a much wider range of temperature and the hysteresis remains prominent as observed in Figure 6 (b). From S#1, S#2 and S#3, $T_h$ decreases with increase in Gd concentration hence the normal state resistivity and this is in accordance with a recently reported work [24]. Thus, $T_h$ can be tuned with the resistivity which can be altered with impurity concentration.

The temperature variations in the critical current ($I_{c0}^{up}$) and switching currents ($I_{c1}^{up}$, $I_{c2}^{up}$ and $I_{c3}^{up}$) for up sweep are displayed in Figure 6(d). Switching currents $I_{c1}^{up}$, $I_{c2}^{up}$ and $I_{c3}^{up}$ related to major voltage steps are denoted by the arrows in Figure 6(a). Above 2.5 K, $I_{c1}^{up}(T)$ strongly deviates and fluctuates from $I_{c0}^{up}(T)$. The region between these two curves (black and green points in Figure 6 (c) / (d)) deals mainly with the flux flow of Abrikosov vortices[10]. While, the region between $I_{c1}^{up}(T)$ and $I_{c2}^{up}(T)$ gets widened with increasing temperature in the range of 2 K to 2.5 K. We can imagine from Figure 6(d) that the extrapolations from these three switching currents towards the lower temperature would converge at some temperature much lower than 2K (not shown in the figure) and below this temperature no voltage steps are expected to appear in the IVCs as no such well-defined steps appear below 6.0 K for S#3, with lower *c*. The appearance of voltage steps in IVCs at relatively higher temperature indicates that they are originated from thermally activated phase slip processes. Now in Figure 6(d), above 2.8K, $I_{c1}^{up}$ is strongly enhanced. At 2.9 K, $I_{c1}^{up}$ follows closely to $I_{c2}^{up}$ and seems to originate from $I_{c2}^{up}$ as it is shown by the dotted connecting line of $I_{c2}^{up}$ and $I_{c1}^{up}$. Similarly at 3.0K, $I_{c1}^{up}$, and $I_{c2}^{up}$, can be linked to $I_{c3}^{up}$ as its origin. This is obvious since at higher temperature the zone close to the superconducting state disappears and the IVC appears as a smooth variation in voltage with current. Thus, the first switching occurs at the next voltage step as it is evident in Figure 6(b).

From the measured IVCs for S#5, we illustrate different regions, bounded by the temperature dependent critical current and switching currents, with colours in Figure 6(d) with respect to their specific roles in controlling the whole process of superconducting-to-normal state transition. With increasing current, the system stays at SC state until the current reaches to the critical current $I_{c0}$ and this state is marked as 'SC' with light green below the $I_{c0}^{up}(T)$. Above $I_{c0}$, the dissipation starts to appear and the mechanism can be characterized mainly either by slow-moving v-av pairs or fast-moving PSLs[13]. The region bounded by $I_{c0}^{up}(T)$ and $I_{c1}^{up}(T)$ mainly shows a smooth variation in voltage with current due to the motion of v-av pairs from the sample edges and the region is highlighted with light pink. Further increasing the bias current voltage jumps to the next resistive state and this can be attributed to the PSLs. The yellow region between $I_{c1}^{up}(T)$ and $I_{c3}^{up}(T)$ indicates that the appearance of resistive states are mainly due to the PSLs. Above $I_{c3}^{up}$, the voltage follows the normal-state (NM) behavior shown by blue part.

## Discussion:

We first emphasize the distinguished observations obtained from the transport measurements performed on NbGd CFs and their distinctiveness from only Nb films. (i) $T_c$ and the transition width strongly depend on Gd concentration and the highest $T_c$ follows from pure Nb film. (ii) IVCs from NbGd samples get equipped with steps-like features which are absent in Nb devices. Increasing slope for subsequent voltage steps and their convergence to the excess current address the origin of these features to phase slip events[10,11,28] stimulated by the magnetic impurities present in NbGd films. (iii) The manifestation of slow moving v-av pairs become more pronounced with increasing Gd and for only Nb film where we observe a single step sharp SC-NM phase transition. (iv) $I_c(T)$, the temperature dependent critical current, strongly depends on Gd concentration while $I_r(T)$, irrespective of Gd amount present in CFs, is mainly controlled by Joule heating across the WLs in the resistive state. $I_c(T)$ and $I_r(T)$ are discussed in detail as follows.

Under the influence of current bias at zero external field, $I_c$ and $I_r$ participate crucially to the SC-NM phase transition. Let us consider different mechanisms governing two important characteristic currents, namely, the critical current $I_c$ and the retrapping current $I_r$. $I_c$ is controlled by the superconducting properties of WLs and $I_r$ deals with the heat dissipation in the resistive state[24]. The commonly used model to describe $I_r(T)$ for hysteretic IVCs is the self-heating model proposed by Skocpol, Beasley, and Tinkham (SBT) [25] which predicts a $(1-T/T_c)^{1/2}$ dependence of $I_r$ on T. On the other hand, the critical current is determined by pair breaking effects in WLs and the standard Ginzburg-Landau (GL) [30] (1-

$T/Tc)^{3/2}$ temperature dependence of $I_c(T)$ can be used for the case of clean superconducting case. However, additional inhomogeneities are known to increase the exponent beyond $3/2$[31].

Combining SBT[25] and GL[30] models, we employ the following equation to inspect the influence of magnetic impurity on $I_c(T)$ and $I_r(T)$:

$$I_{c,r} \alpha (1-t)^s \; ; \; t = T/Tc \qquad (1)$$

Where the exponent *s* combines both $I_r$ and $I_c$ in a single equation and t ($t = T/Tc$) being the reduced temperature. $s = 1/2$ for $I_r$ indicates the self-heating as the origin of hysteresis[25] while, $s = 3/2$ for $I_c(T)$ represents the GL theory of depairing current limit [30]. In Figure 7, $I_r(T)$ (top panel) and $I_c(T)$ (bottom panel ; $I_c = I_{c0}$) are displayed for one Nb sample (S#1) and two NbGd samples (S#3 and S#5).

The red solid line, in Figure 7(a) exhibiting $I_r(T)$ for the Nb device S#1, is the best fit using equation (1) with $s = 0.6 \pm 0.01$, while the black dashed line represents the fit using SBT model. As the dashed line appears very close to the experimental points and the difference in the *s* values for the red line and the dashed line is not much, the $I_r(T)$ and hence the hysteresis can be described by self-heating. For $I_c(T)$ of Nb device in Figure 7(b), the best fit provides an exponent of unity which is consistent with the WLs based Josephson Junction near $T_c$ [32]. $I_r(T)$ and $I_c(T)$ For NbGd Sample S#3 have been shown in Figures 7(c) and 7(d), respectively. $I_r(T)$ clearly follows the self-heating model, i.e., $I_r \alpha (1-t)^{1/2}$. The exponent for the best fit of $I_c(T)$ appears as 1.69 which is higher than the exponents for pure Nb and the depairing limit. Similarly, Figures 7(e) and 7(f) exhibit the results of $I_r(T)$ and $I_c(T)$ for sample S#5, respectively. A clear deviation is observed for $I_r$ from the self-heating model (black dashed line) to the best fit (solid line with *s* = $0.66 \pm 0.02$). However, the value of the exponent is little higher than that for the Nb sample and 0.5 the exponent from the SBT model. This deviation could arise from the $T_c$ determination since the transition is broad and is equipped with multiple steps hence the exact retrapping may occur at much lower current. The $T_c$ is determined from its IVC. A very large value of *s* ($s = 3.12 \pm 0.18$) appearing in the best fit of $I_c(T)$ clearly dictates the pronounced effect of Gd concentration on $I_c(T)$. A larger exponent, as observed for high-$T_c$ SC films embedded with FM top layer[31], indicates a stronger decay in the critical current with temperature and reveals the possibility of multiband superconductivity with band coupling [33].

Finally, the results strongly support that a very little amount of ferromagnetic inclusion in SC-FM based CFs leads to the initiation and modulation of phase slip processes. SC-FM based CFs with dilute magnetic doping can provide a platform to study the varieties of physical phenomena like creation and annihilation of v-av pairs along with their motion under electric field, initiation and manipulation of phase slip processes etc. The variation in $T_c$ and other critical parameters like, critical field, critical current, transition width etc. with Gd concentration will be discussed in detail elsewhere. Further, above $T_h$, the hysteresis in IVCs does not disappear and $I_r$ becomes larger than $I_c$ in NbGd devices. A more detailed investigation should be pursued experimentally as well as theoretically which include the contribution from the magnetic part to the SC-NM transition. The influence on magnetic component on $I_c(T)$ is noticeable and further studies needed to explore the possibility of multicomponent superconductivity with interband coupling between two metals[33]. Moreover, when decreasing the geometrical dimension becomes technologically difficult and challenging, the presented method can be useful for one step creation of nanometer-scaled weak links in a superconductor and hence can be a promising candidate towards the fabrication of phase slip based devices for future photonic and metrological applications.

# Methods:

The CFs were grown in an ultra-high vacuum (UHV) dc magnetron sputtering system by co-sputtering of Gd (99.95%) and Nb (99.99%) directly on Si (100) substrate having 300 nm thick thermally oxidized $SiO_2$ as the dielectric spacer between the substrate and the film. Before the deposition run the sputtering chamber was evacuated to less than 5 x $10^{-9}$ Torr and the co-sputtering was performed in an Ar (99.9999% purity) environment at about 3

mBar. Nb and Gd were co-sputtered with rates 3 Å/sec and 1-2 Å/sec respectively. With optimized position of the samples and controlled opening of the shutter for Gd target we fabricated CFs mainly with varying Gd concentration in one sputtering run while keeping other parameters unchanged. Even for different deposition runs the relevant parameters were maintained very closely so that the results could be comparable from samples fabricated in different sputtering runs.

Devices were fabricated using shadow mask for conventional 4-terminal measurements. Current and voltage leads were defined with Au (100nm)/Ti (20nm) layers followed by the fabrication of NbGd CFs based channel of width about 100 micron. A Si capping layer of about 10 nm thickness was finally sputtered on the CF to avoid any oxidation while exposed to the atmosphere. Total channel length between the voltage leads was 1300 micron for all the devices for which the results reported here. Careful measurements were done for the estimation of the concentration by energy dispersive spectroscopy (EDS) analysis using a field emission scanning electron microscope by Zeiss. For each device, we determined the composition by EDS at minimum 5 places and the average value was considered for the analysis within an accuracy of 0.20 at. %. The morphological studies were performed by an atomic force microscopy (AFM) in tapping mode. The transport measurements were carried out in a Physical Properties Measurement System (PPMS) equipped with 14 T magnet by Quantum Design. The devices were mounted on a puck and soldered for the electrical measurements.

# Acknowledgments:

We thank prof. R.C. Budhani for the fruitful discussions. We are indebted to Dr. V. P. S. Awana for his support in using PPMS system to carry out this work and also for critical reading of the manuscript. We gratefully acknowledge Dr. Sudhir Husale for the invaluable discussions and for reviewing the manuscript. The technical help for AFM and FESEM imaging facilities in CSIR-NPL are highly acknowledged. We are thankful to Mr. M. B. Chhetri for his technical help. A. B. acknowledges CSIR-NPL for providing the research internship and R. J. acknowledges CSIR for senior research fellowship. The work was supported by CSIR network project 'AQuaRIUS' (Project No.: PSC 0110).

# Author contributions:

A.B. fabricated the devices and participated in the transport measurements. R.J. participated in the transport measurements. S.S. planned the project, designed and fabricated the devices, performed experiments, analyzed and interpreted the data and wrote the manuscript. All the authors read and reviewed the manuscript.

## Additional information

Competing financial interests: The authors declare no competing financial interests.

## Figure Captions:

**Figure-1: Surface morphology of CFs.** Atomic force microscopy (AFM) image of 100 nm thick Nb thin film deposited on $SiO_2$/Si substrate at room temperature by UHV magnetron sputtering.

**Figure 2: Resistivity-Temperature (R-T) characteristics of Nb and NbGd devices.** (a) Linear scale representation of R-T measurements for one Nb and 4 NbGd devices with varying Gd concentration. Inset: SEM image of a typical device geometry with current and voltage geometry. (b) Semi-logarithm plot of the same data shown in (a).

**Figure 3: Effect of magnetic field on resistance.** (a) R-T curves for the sample S#5 at different magnetic field applied perpendicular to the plane of the film.(b) Magnetoresistance for the same sample

(S#5) measured at different temperature. Dashed arrows in both (a) and (b) indicate existence of a plateau region in resistance closed to zero resistance state.

**Figure4: Current-voltage Characteristics (IVCs) of a Nb sample (S#1).** (a) IVCs at different temperature for Up and Down sweeps as indicated by the arrows. (b) Representation of the same IVs shown in (a) in a 3-D format where individual IVC isotherm is separated along z-axis. Inset: Dependence of critical currents and re-trapping current with temperature. Details about the current values are shown in (b) with arrows and are discussed in the text.

**Figure 5: IVCs of the NbGd sample, (S#3).** (a) Zero-field isothermal IVCs for up and down sweeps in anti-clockwise directions as indicated by the arrows. (b) Selective representation of four IV isotherms taken from (a). The IVCs, measured at 6.1 K, 6.2 K and 6.3 K, have been shifted upward by 50 mV, 100 mV, and 150 mV, respectively, for clarity. Solid (open) symbols represent up (down) sweep direction. Dotted lines converge to a single point measuring the excess current. (c) Dependence of critical currents and re-trapping current with temperature. Details about the determination of current values are shown in (b) with arrows.

**Figure 6: IVCs for NbGd sample, S# 5** (a) A set of isothermal IVCs for up and down sweep directions measured at zero external field. b) IVC isotherms shown in (a) are plotted separately with an upward voltage shift of steps 35 mV from 2.1K up to 3.0K with 0.1 K intervals in temperature respectively, for clarity. Solid lines (open symbols) represent up (down) sweep directions. The hysteretic regions for 1st, 2nd and 3rd step switching are shown by dotted lines along with its numbering. (c)Variation of $I_{c0}^{up}$, $I_{c1}^{up}$, $I_{c1}^{dn}$, $I_r$ with temperature. (d) Representation of temperature dependent critical current and three major

switching currents defined by the featured voltage steps appeared in the IVCs for up sweep direction. Various characteristics current values are defined in (a) and (b) with arrows. Different zones, highlighted with the colors in the current-temperature plot, are defined with respect to their electronic properties and related to dissipation mechanisms.

**Figure7: Variation of $I_{c0}$ and $I_r$ with reduced temperature, t ($t=T/T_c$), for samples S#1, S#3, and S#5.** Dependence of $I_r$ with t for (a) Nb Sample S#1, (c) NbGd sample S#3, and (e) NbGd sample S#5. The critical current, $I_c$ as a function of reduced temperature for (b) Nb Sample S#1, (d) NbGd sample S#3, and (f) NbGd sample S#5. Black open scattering points represent the experimental data, red solid lines represent the best fit using equation; $I_{c,r} \propto (1-t)^s$, where the values for s are indicated in individual plots. Dashed lines in (a) and (e) represent the fit with s = 0.5 corresponding to the self-heating model. The details are explained in the text.

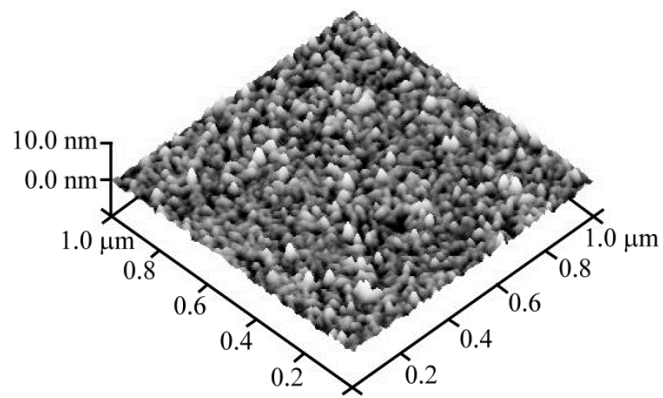

Figure-1

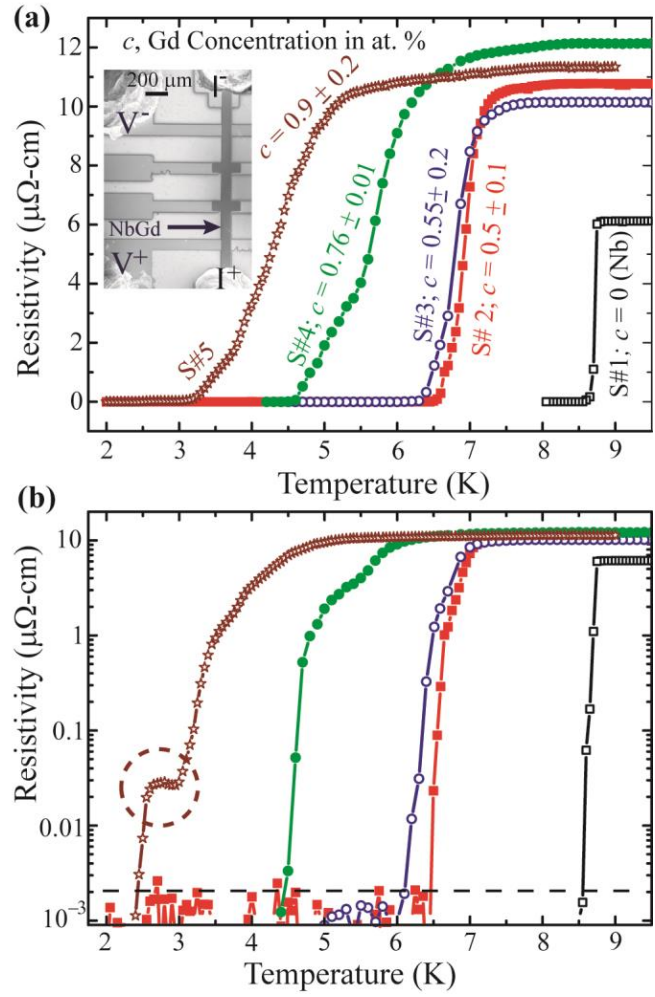

Figure-2

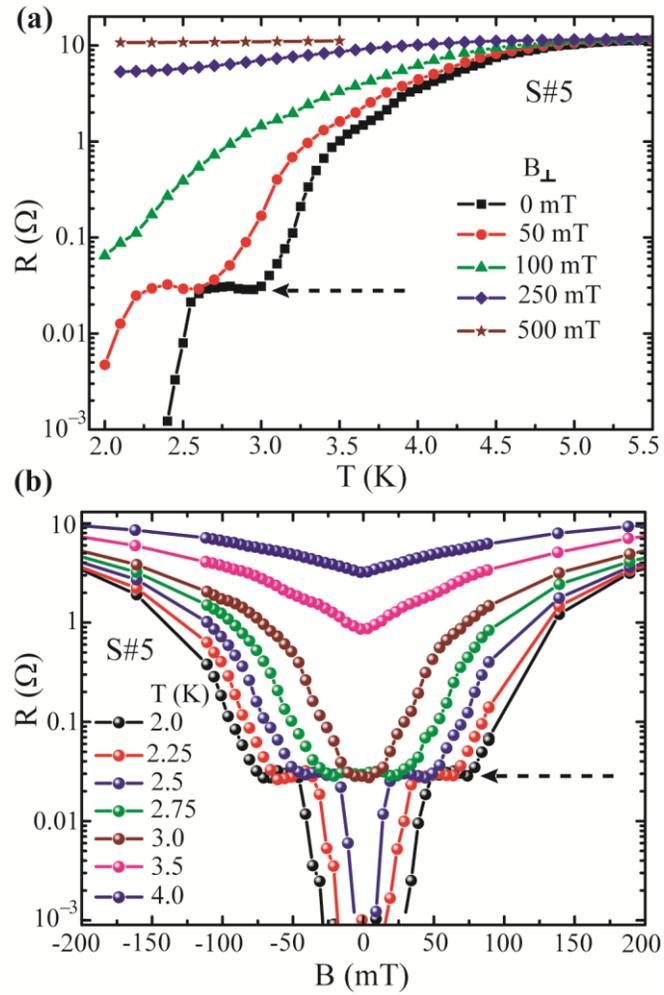

Figure-3

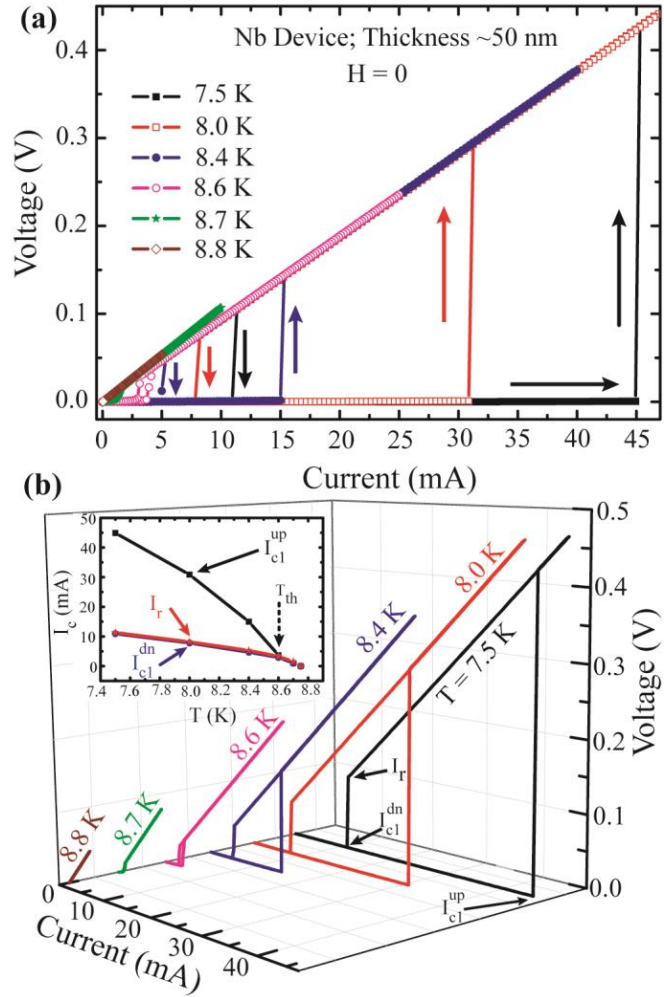

Figure-4

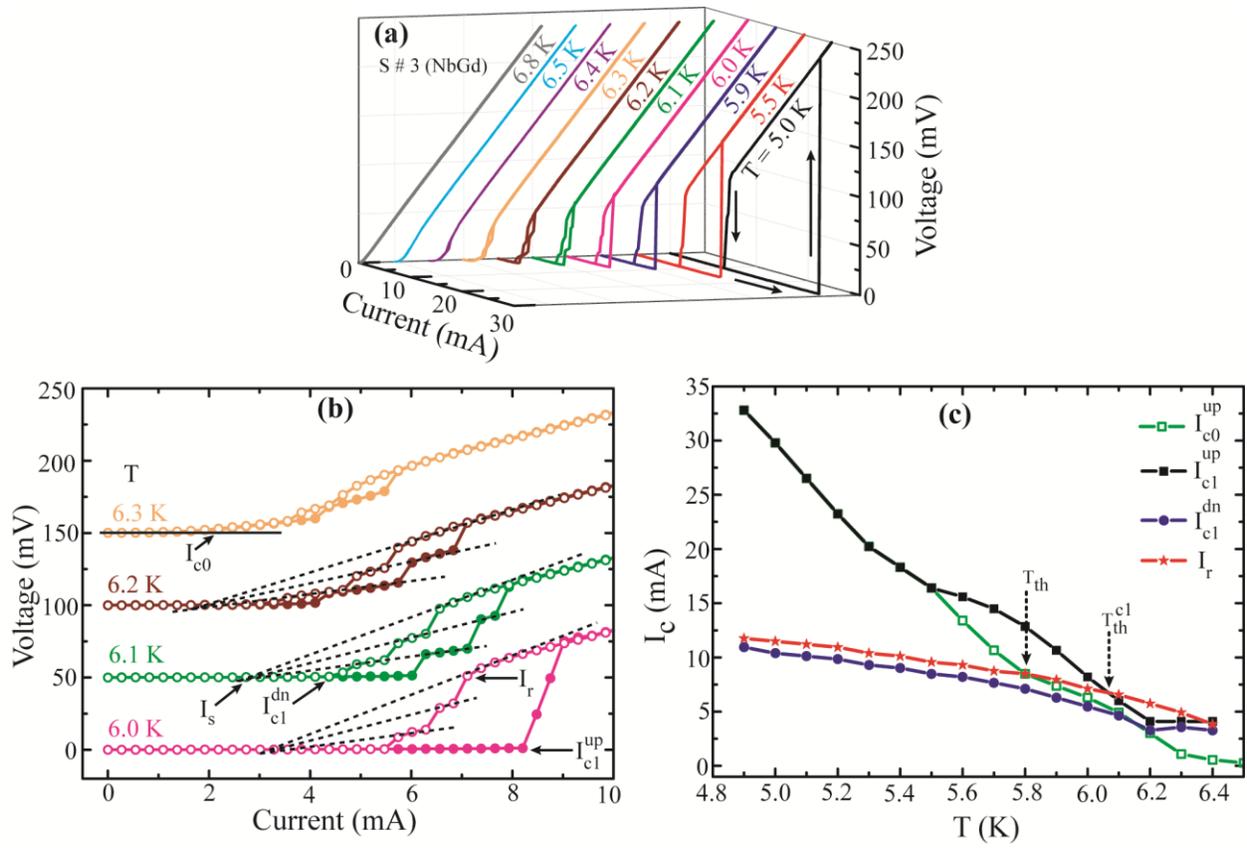

Figure-5

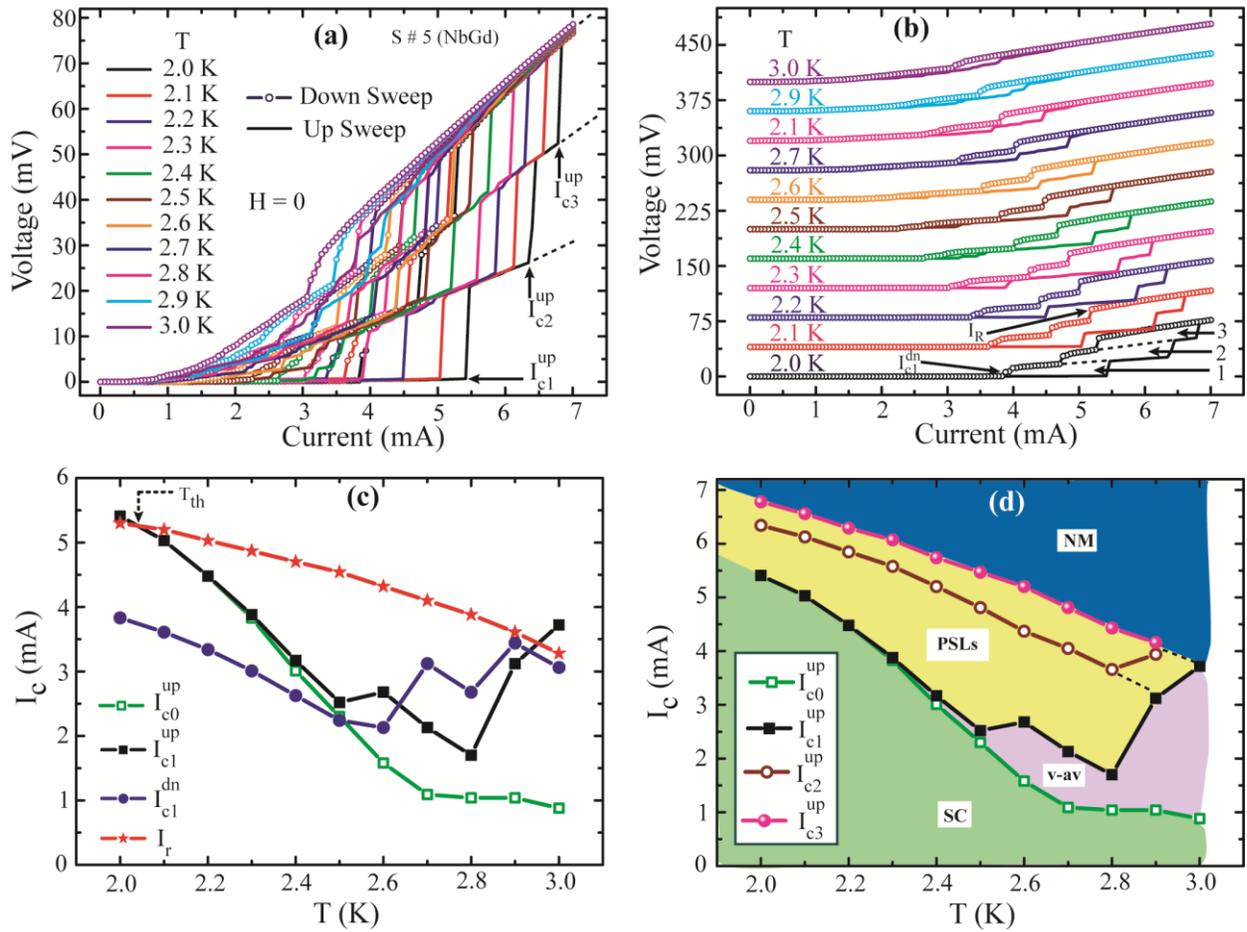

Figure-6

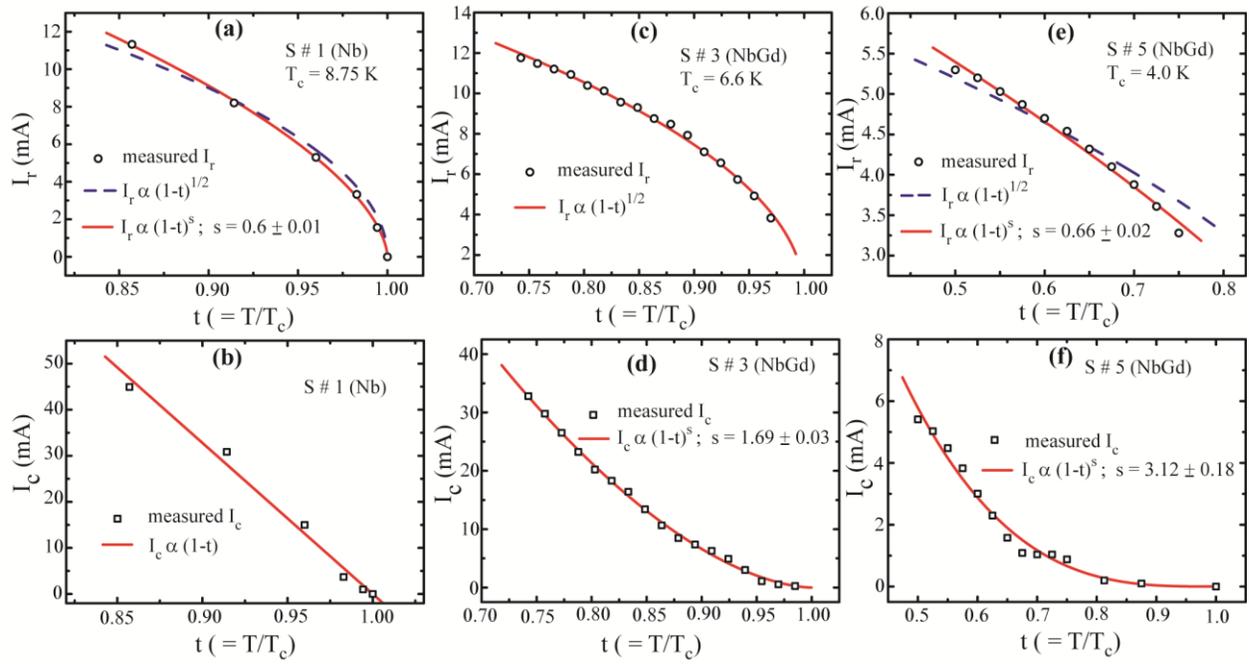

Figure-7